\documentclass[12pt,preprint]{aastex}

\begin{document}

\def\chandra{{\em Chandra}}
\def\e05{0509$-$67.5}

\title{Raising the Dead: Clues to Type Ia Supernova Physics from the Remnant 0509$-$67.5}

\author{Jessica S. Warren and John P. Hughes}
\affil{Department of Physics and Astronomy, Rutgers University}
\affil{136 Frelinghuysen Road, Piscataway, NJ 08854-8019}
\email{jesawyer@physics.rutgers.edu, jph@physics.rutgers.edu}

\begin{abstract}

We present \chandra\ X-ray observations of the young supernova remnant
(SNR) \e05 in the Large Magellanic Cloud (LMC), believed to be
the product of a Type Ia supernova (SN Ia).  The remnant is very round
in shape, with a distinct clumpy shell-like structure that extends to
an average radius of $14.8\arcsec$ (3.6 pc) in the X-ray band.  Our
\chandra\ data reveal the remnant to be rich in silicon, sulfur, and
iron.  The yields of our fits to the global spectrum confirm that \e05
is the remnant of an SN Ia and show a clear preference for delayed
detonation explosion models for SNe Ia.  The \chandra\ spectra
extracted from radial rings are in general quite similar; the most
significant variation with radius is a drop in the equivalent widths
of the strong emission lines right at the edge of the remnant.
We study the spectrum of the single brightest isolated knot in the
remnant and find that it is enhanced in iron by a factor of roughly
two relative to the global remnant abundances. This feature, along
with similar knots seen in Tycho's SNR, argues for the presence of
modest small-scale composition inhomogeneities in SNe Ia.  The
presence of both Si and Fe, with abundance ratios that vary from knot
to knot, indicates that these came from the transition region between
the Si- and Fe-rich zones in the exploded star, possibly as a result
of energy input to the ejecta at late times due to the radioactive
decay of $^{56}$Ni and $^{56}$Co.
Two cases for the continuum emission from the global spectrum were
modeled: one where the continuum is dominated by hydrogen thermal
bremsstrahlung radiation; another where the continuum arises from
non-thermal synchrotron radiation.
The former case requires a relatively large value for the ambient
density ($\sim$1 cm$^{-3}$).  Another estimate of the ambient density
comes from using the shell structure of the remnant in the context of
dynamical models.  This requires a much lower value for the density
($<$0.05 cm$^{-3}$) which is more consistent with other evidence known
about \e05.  We therefore conclude that the bulk of the continuum
emission from \e05 has a non-thermal origin.

\end{abstract}

\section{INTRODUCTION}

Type Ia supernovae (SN Ia) are characterized by the absence of Balmer
lines and the presence of strong silicon lines in their optical
spectra \citep{hillnie}.  They yield mostly intermediate mass elements
from Si to Ca and are the main sources of iron production in galaxies.
Each SN Ia produces $\sim$0.8 ${\rm M{_\sun}}$ of Fe \citep{iwam},
making them important for understanding the chemical evolution of
galaxies.  The peak magnitude and shape of their light curves are
fairly homogeneous, which makes them useful as distance indicators and
tools for cosmology \citep{rpk,riess,perl}.  For these reasons, SN
Ia's are the objects of much study.

Still, there is much that remains uncertain about them.  The
progenitors of SN Ia's are believed to be degenerate C+O white dwarfs.
Although models for sub-Chandrasekhar mass explosions exist, the
favored explanation is that the degenerate object somehow exceeds the
Chandrasekhar limit before exploding.  Two possibilities include
a single degenerate scenario, in which the white dwarf accretes mass
from a companion star, and a double degenerate scenario, in which a
pair of white dwarfs coalesce \citep{branch}.  It is still unclear
whether one, both, or none of these offers the correct explanation.  In
addition, the mechanism of the explosion, i.e., where the ignition
begins and how it propagates outward, continues to elude us
\citep{iwam}.  Two cases studied in detail include a fast deflagration
and a delayed detonation.  In the former, a deflagration begins close
to the center of the progenitor star and propagates subsonically.  The
delayed detonation case involves a deflagration initially, which is
transformed at some density to a detonation as the star expands.
Models for these two cases are parameterized by the speed of the
burning front and the location of the ignition in density space;
we compare our results with those of \citet{iwam}.

By studying the remnants of SN Ia explosions, we can hope to gather
clues to some of these puzzles.  The supernova remnant (SNR) \e05 in
the Large Magellanic Cloud (LMC) is believed to be the remnant of an
SN Ia \citep{tuohy, hughes95}.  It was discovered in an X-ray survey
of the LMC with the {\it Einstein X-ray Observatory} \citep{long}.
However, much of the previous study on this remnant was done via
optical spectroscopy.  In H$\alpha$, the remnant appears as a very
circular shell \citep{tuohy}.  \e05 shows no detectable emission in
[\ion{O}{3}] and [\ion{S}{2}], indicating it is a Balmer-dominated
remnant \citep{tuohy, smith91,smith94}.  Remnants such as this
generally have two components to the H$\alpha$ emission line: a narrow
component produced by neutral hydrogen being excited by the shock, and
a broad component produced by charge exchange with fast protons
post-shock.  In \e05, Smith et al. (1994) found the narrow components
to be $\sim$25--31 km/s, higher than expected for neutral hydrogen
and a possible indication of the presence of a cosmic ray precursor.
This SNR is also intriguing in that its broad component is believed to
be so broad that it has escaped detection, pointing to very fast
shocks ($\geq$2000 km/s) and a young age ($\leq$1000 yr) (Smith et
al. 1991; Smith et al. 1994).  \citet{tuohy} found that \e05 is
expanding into a low density medium, with $n_{\rm H}\lesssim0.02$
cm$^{-3}$.  A previous X-ray study of this SNR with {\it ASCA}
revealed it to be strong in Si and Fe L emission, with little O, Ne,
and Mg emission \citep{hughes95}.  This remnant has also been observed
in the radio, showing a spectral index of $\alpha=0.46$
\citep{mathewson} and a flux density of 0.066 Jy at 1 GHz
\citep{henrey}.

\section{OBSERVATIONS}

\e05 was observed with the \chandra\ {\em X-ray Observatory} for 51.4
ks on 2000 May 12-13 on the back-side--illuminated chip (S3) of the
Advanced CCD Imaging Spectrometer (ACIS-S) (Obs ID 776).  We made use
of the charge transfer inefficiency (CTI) corrector code
\citep{townsley} to correct for spatial gain variations.  Standard
software tools were used to filter for grade (retaining the usual
grades 02346 only), bad pixels, and times of high or flaring
background.  After all filtering, the net exposure time was 47.9 ks.
All spectral analyses utilized the pulse-invariant (PI) column of the
final \chandra\ events file.

As described in \S 4.1.1 as well as in \citet{hughes95}, the
spectrum of \e05 includes a strong Si line.  We measured the centroid
of this Si K$\alpha$ line to be low, $1.835\pm0.002$ keV, which is
indicative of a very low ionization state for Si, i.e., Li- or Be-like.
Most SNRs generally show either the H-like Si K$\alpha$ transition at
2 keV or the He-like Si K$\alpha$ complex at $\sim$1.86
keV\footnote{Values obtained from the CXC Atomic Database:
http://obsvis.harvard.edu/WebGUIDE/}.

\chandra\ has onboard radioactive sources for calibration, which we
used to independently verify the energy scale of the detector at the
time of the \e05 observation.  Two calibration observations (Obs IDs
62057 and 62058) were taken just prior to ours, and two more (62055
and 62056) were taken beginning the day after.  After running the CTI
corrector code on these data, we fit gaussian lines to the spectra of
the sources.  The most prominent lines were Al K$\alpha$, Ti
K$\alpha$, Mn K$\alpha$, and Mn K$\beta$.  The centroids of these
lines were compared with the accepted values \citep{crc} and found to
have less than 0.2\% difference.  Most of the centroid values agree
with the accepted values within the errors (see Table 1).  The
residual uncertainty in the energy scale is of order a few eV at both
$\sim$1.5 keV and $\sim$6.5 keV, which is the order of the uncertainty
in our Si line centroid.  Therefore, we can be confident that we do
indeed have a low ionization state for Si.

\section{IMAGING}
\subsection{\chandra\ Image}

In Figure 1 we present the 3-color \chandra\ broadband (0.2--7 keV)
X-ray image of \e05.  Red contains emission from 0.2--0.69 keV, green
0.69--1.5 keV, and blue 1.5--7 keV.  The image reveals a clear shell
of emission which is very circular, with a slight break in the
northwest.  The shell itself appears clumpy, with small knots of
emission spaced seemingly periodically around the bright rim.  In
fact, most of the structure in this remnant can be characterized as
clumpy, as there are no filaments or elongated features visible.  On
the northern portion of the rim and more faintly on the southern
portion, the higher energy band emission is slightly enhanced.  In
the southeast, we see some enhancement in the lower energy band.  The
rest of the rim shows less spectral variation.  Most of the interior
shows only faint emission, except for an enhanced clumpy region in the
southwest and a lone knot in the north, both clearly dominated by the
middle energy range, where Fe L emission is strong.  There seem to be two
distinct halves to the SNR, the dividing line running at an angle of
$\sim$30$\degr$ west of north.

Separating the different energy bands, as presented in Figure 2,
reveals essentially the same subtle variations as noted in the color
image.  Here we chose narrower spectral bands that contain emission
mostly from O (0.45--0.7 keV), Fe L (0.7--1.4 keV), and Si (1.5--2
keV).  In the O band, the knot in the north and the clumpy region in
the southwest are much fainter.  The rim here is less apparent than in
the other bands, although there does seem to be a slight enhancement
in the southeast, present in the color image as well.  The brightness
of the interior emission is also more uniform here.  As before, the Fe
L band reveals the clumpy region and knot to be quite prominent, and it
is in this energy band that these spatial features are most obvious.
The Si band shows fainter emission in the interior.
The knot and clumpy region are apparent here, and the northeastern rim
appears slightly brighter at these energies.  Spectral differences
within the remnant are small and are mostly associated with the knot,
clumpy region, and outermost edge (see \S 4.2).  The CTIO 4-m
H$\alpha$ image \citep{smith91} also reveals a shell of emission, with
slight enhancements in the north and southwest, and a fainter area in
the southeast.  There is virtually no H$\alpha$ in the interior.

\subsection{Image Fits}
The distinct shell-like nature of the remnant prompted us to
investigate this structure further.  We did this in two ways: fitting
a simple geometric model to the image, and performing a deprojection
of the normalizations derived from spatially-resolved spectra, which
is described in \S 5.  Our specific goals for the simple image
fits were modest: (1) to determine the inner and outer radii of the
shell of X-ray emitting material, and (2) to estimate the total SNR
X-ray flux that comes from clearly resolved bright knots.  We used a
model consisting of two nested elliptical shells and a number of
spherical knots that describe the observed X-ray surface brightness
of the SNR. Each individual component was assumed to be optically thin
and have uniform emissivity, so that the projected surface brightness
distribution was uniquely determined by the geometry of the model.
The shell models shared the same center, axial ratio, and position
angle for the major axis, assumed to lie in the plane of the sky. The
inner edge of the outer shell and the outer edge of the inner shell
were set equal (i.e., the shells were adjoining).  Thus, in total
there were nine parameters for the shell models: the center position
in R.A. and declination, axial ratio, position angle, outer radius of
the outer shell, common radius, inner radius of the inner shell, and
the intrinsic emissivities of the outer and inner shells.  Each
spherical knot required four parameters: center position in R.A. and
declination, radius, and intrinsic emissivity. The model surface
brightness was convolved with the \chandra\ point spread function
(PSF) appropriate to the center of the SNR at a monochromatic energy
of 0.75 keV (where the observed spectrum peaks).  The PSF was
determined using the \chandra\ Ray Tracing (ChaRT) program (described
in \S 5).  Only the symmetric northeastern portion of the SNR was
fit to avoid complications with the clumpy region just inside the
main shell visible on the southwestern side.

The software we used is described in \citet{hughbirk}.  It is based on
a maximum likelihood figure of merit function derived for
Poisson-distributed data and allows for the determination of both
best-fit parameter values as well as confidence intervals.  We employ
the robust downhill simplex method \citep{press} for function
minimization.

Fits were done in an iterative fashion, starting with the shell models
alone and then adding successively more and more spherical knots, as
needed.  Knot models were introduced at the locations of the most
significant residuals between the data and the model from the previous
iteration. Iterations were continued until the absolute value of the
residuals fell below $\sim$10 X-ray events/pixel. Knots representing regions
of excess emission (positive residuals) were modeled with positive
intrinsic emissivity values.  We also allowed for negative emissivity
knots to account for negative residuals, i.e., regions of lower
overall emission.  The physical interpretation is that these knots
represent places in the shell with lower than average density or
thickness that appear faint relative to their surroundings.  In models
with such negative knots, the overall modeled surface brightness was
required to be non-negative everywhere before PSF convolution in order
to ensure a physically realistic model.

The shell model is intended to describe the overall limb-brightened,
global structure of the remnant, and the knots represent additional intensity
fluctuations on spatial scales of an arcsecond or so.  Due to the
simplicity of the geometric structure models and the large number of
arbitrary free parameters (e.g., four for each knot model), our image
fitting procedure does not lead to a unique solution for the structure
of the remnant. In addition we found that the precise manner in which
knot models were introduced resulted in noticeable differences in the
derived remnant structure.  However, our intent here is not to derive
a unique spatial model, but rather one that is broadly representative
of the spatial distribution of X-ray flux from \e05. The shell and
spherical knot models provide a convenient way to parameterize this.
We experimented with different methods for introducing models during
the iterative fitting process and found that our quantities of
interest (the inner and outer radii of the shell model and the average
ratio of number of X-ray events to volume for the knots) were fairly 
robust against variations in the fitting procedure. 

Figure 3 shows the results of the image fits. One measure of the
success of our image fitting procedure is the good agreement between
the ``Data'' and ``Model'' panels in the figure.  In this best-fit
model there are 28 knots of positive emission and 5 knots of negative
emission. Most of the knots are effectively unresolved (i.e., fitted
sizes less than or comparable to the \chandra\ PSF) and appear to be
distributed non-uniformly over the portion of the image
fitted. However, the radial distribution of knots is consistent with
that expected for a uniform distribution within the volume of a thin
spherical shell.  Comparison of the FFT power spectra of the raw data
image and best-fit model reveal the presence of additional significant
power in the data, above Poisson noise, on the smallest spatial
scales. Further support for this is evident in the ``Difference''
panel of Figure 3, which we interpret as showing intensity
fluctuations on smaller spatial scales. Thus our image fits have only
captured the brightest and most obvious small-scale structure in the
image and what we have modeled as diffuse shell-like emission actually
consists of knots extending to scales below our image resolution. This
is most likely similar to the ``fleecey'' structure evident in
\chandra\ images of Tycho's SNR, which may be a typical feature of SN
Ia remnants.

There were 38,900 X-ray events in the fitted portion of the image.
The fit yielded 11,100 (21,600) in the outer (inner) shell, 6800 in
positive knots, $-$900 in negative knots and roughly 100 in
background. (The values do not sum due to rounding.)  Of most interest
is the relationship between the number of detected X-ray events and
the geometric volume of the component.  In particular we want to
compare the number of events and the volume from the knot component to
that of the shell component.  Using the above values, the ratio of
X-ray events in the positive knots to the shells is $\sim$20\%.  The
volume corresponding to the sum of all the fitted knots is only 2\% of
the total volume of the shell component.  We use these values in \S6.4
below.

The inner and outer radii (more precisely the semi-major axis lengths)
of the shells determined from our fits are 15.2$\arcsec$ and
12.8$\arcsec$, which correspond to physical radii of 3.69 pc
and 3.10 pc for an LMC distance of 50 kpc (assumed throughout). The
axial ratio for the elliptical shells is 0.945 and the major axis lies
at a position angle of 7.6$\degr$ west of north.  The outer edge of
the X-ray emitting shell corresponds nearly perfectly with the outer
edge of the H$\alpha$ image of \e05 from \citet{smith91}.  This
indicates that the outer edge of the X-ray emission corresponds to the
location of the blast wave propagating into the interstellar medium.
We suggest that the inner edge of the X-ray emitting shell corresponds
to the location of the reverse shock propagating backward into the
ejecta.

\section{SPECTRA}
\subsection{Global Spectrum}
We describe the model fits to the global, or integrated, spectrum of
\e05 (see Figure 4) in some detail.  The global spectrum provides us
with the elemental yields for this remnant, which we will use to argue
for a Type Ia origin (see \S 6.1).  In addition, the best-fit
models for this spectrum are used as templates for the spectra of
half-ring regions used to determine the shell structure of \e05 (see
\S 4.2).

\subsubsection{Basic Model Fits}
It was immediately apparent that \e05 was rich in Si, as evidenced by
a strong Si K$\alpha$ emission line in the global spectrum (see Figure
4).  Because of this, we attempted to fit the high end ($>$1.6 keV)
portion of the spectrum and then extrapolate back for the lower
energies utilizing the minimum number of free parameters necessary to
obtain a good fit.  This regime includes K-shell lines of Si, S, Ar,
Ca, and Fe, while excluding the strong Fe L-shell emission, for which
the atomic physics is not as well understood.  We used a
non-equilibrium ionization (NEI) plane-parallel shock model
\citep{hughes00} to do the fitting, allowing column density,
temperature, terminal value of the ionization timescale, and the
elemental abundances of Si, S, Ar, Ca and Fe to vary freely.
Abundances are relative to the solar values of \citet{ag}.  Hydrogen
was used to model the continuum, and no other species were included.
LMC abundances of 0.3 times solar were used to determine the column
density, and an additional Galactic column density of
$5.5\times10^{20}$ cm$^{-2}$ \citep{dl} was incorporated at solar
abundances.  The spectrum was background subtracted using an annular
region surrounding the SNR.

We obtained good fits to the high energy portion of the spectrum with
abundances given in Table 2, as long as Si was allowed a slightly
lower ionization timescale than the other elements, and Fe was allowed
both a higher temperature and higher ionization timescale.  However,
this particular fit overpredicted the emission below $\sim$ 1 keV.  In
order to fit the lower energies, we added in O, Ne, and Mg. We then
stepped through the temperature until we found a good fit.  The
temperature increased from the best value of the high energy fit,
while the ionization timescales decreased (this significantly improved
the fit below 1 keV).  The abundances of Si, S, and Ar decreased only
slightly; Ca showed a more pronounced decrease, and Fe went up (see
Table 2).  While the fits definitely required O and Mg emission,
adding Ne resulted in no improvement. There was a need for a line at
2.11 keV that was not included in the NEI model.  We believe this to
be a K$\beta$ line of Si, which we fit with a gaussian line.

There was also a portion of the spectrum around 0.73 keV, presumably
Fe L-shell emission, which the model could not fit.  It was possible
to compensate for this deficiency by including a single narrow
gaussian line with freely varying energy centroid and intensity.
Because this is the brightest part of the spectrum, the $\chi^2$
statistic weights data points in this vicinity more than those at
lower or higher energies, even though our understanding of the atomic
physics here is poorest.  We therefore introduced a systematic error
of 5\% to the data so that our reduced-$\chi^2$ value was 1, and the
statistical weight of points near the Fe L-shell emission was reduced.
An advantage of having a reduced-$\chi^2$ = 1 is that we can derive
errors on fitted quantities in the usual manner.  We should stress
that while there are inadequacies in the model, it still does a good
job of fitting the rest of the Fe L emission which extends from
$\sim$0.8--1.3 keV, as well as the K-shell emission lines.  So, while
the model could be more accurate in some regions, we are confident
that it is good {\em overall}.

\subsubsection{Two Models of Continuum}
Continuum emission in SNRs can come about in various ways.  It could
arise from the metals in the ejecta alone.  However, the equivalent
widths of the line emission we see in \e05 are not as high as we would
expect to see in this case, so another source of continuum emission is
required.  The thermal bremsstrahlung radiation of electrons on
ionized hydrogen, from either the interstellar medium (ISM) or the
ejecta itself, can produce sufficient continuum emission. We modeled
this case with hydrogen as the continuum component (hereafter, Case
H).  We assumed the continuum temperature and the temperatures of the
metals, excepting Fe, were the same.  No He or other species were
added to the continuum model as we were attempting to fit our spectrum
with the least number of physically plausible parameters.

A second case we modeled was for non-thermal X-ray synchrotron
radiation.  SN1006 and other SNRs show approximately power-law
spectra in the X-ray band which is now believed to be the signature of
a population of relativistic particles accelerated by the SN shocks
\citep{koy,allen97}.  It has been observed that this X-ray power-law
spectrum steepens from the power-law at radio frequencies.  Simple
parameterized models have been developed to account for this.  One such
is the synchrotron cut-off (SRCUT) model, that takes as input the
radio flux at 1 GHz and the radio spectral energy index, and fits for the
so-called ``roll-off'' (or ``cut-off'') frequency.  This is the
frequency at which the X-ray flux has fallen to about a factor of 10
below the extrapolated radio spectrum \citep{reykeo}.  The SRCUT model
was applied to \e05 by \citet{henrey} using archival {\em ASCA} and
radio data, and they obtained an upper limit to the roll-off frequency
of 2.9$\times 10^{16}$ Hz.  Using our \chandra\ data, we applied the
SRCUT model in the context of our global fit (hereafter, Case S) and
were able to get good fits to our spectrum.  In addition, our result
for $\nu_{roll-off} = 1.6\times 10^{16}$ Hz falls well below their
limit.  We also fit a power-law model to the spectrum in place of the
SRCUT model to obtain the photon index.  A value of $3.25\pm0.18$ and a
normalization at 1 keV of $2.2\times10^{-7}$ Jy fit the data well.

In Figure 4, we plot the global spectrum of \e05 overlaid with the
best-fit Case H (solid line) and Case S (dotted line) models.  Up to
about 3 keV the two models are virtually indistinguishable.  Above 3
keV, Case H overpredicts the continuum level and Case S fits the data
better. In Table 3 we present the best-fit parameters for Cases H and
S for the global spectrum.  Here, $n_en_iV/$[{\em i}/H]$_\sun$ is the
emission measure (EM) of each species, $i$, relative to the solar
abundance for that species.  This was done so that if we were looking
at a solar plasma, the quantities listed would all have the same
value.  For Case H we can also determine absolute abundances relative
to solar; those values are given in the last column of Table 2. For
Case S this is not possible since we have set the EM of hydrogen to
zero.  Other than that, there are no glaring differences in EMs
between the two cases. In general, Case S shows slightly lower EMs
than Case H.  The temperature for Case S is a bit higher, as well.  As
mentioned in \S 4.1.1, we included a systematic error in our data when
fitting for Case H.  This was carried over into our fits for Case S,
but in fact may be a slight overestimate for this case (since the
overall fit is somewhat better for Case S).

\subsection{Ring Spectra}
Given \e05's very circular, unbroken appearance in both the optical
and X-ray images, we decided to investigate the spectrum as a function
of radius.  The two ``halves'' of the SNR mentioned in \S 3.1 provided
a natural dividing line down the remnant where the brightness jumps.
The northeast hemisphere was divided into six half-rings, each
containing approximately the same number of X-ray events and with a
minimum width of 2 pixels ($\sim$1\arcsec, the resolution limit).
These half-rings were then extended to the southwest hemisphere as
well to have matching regions.  This left us with twelve regions: six
half-rings in each of two hemispheres (see Figure 5).  The rings are
numbered from 1 (inner) to 6 (outer).  We used the best-fit model to
the integrated spectrum for Cases H and S as templates for each of
these regions, allowing only the column density and elemental
abundances for the regions to vary.  No systematic error was
introduced for these spectra.  In all cases we obtained good fits in
this manner, indicating that the principal differences in spectra in
the remnant arise from abundance variations, as the column density
does not vary widely across the remnant.

The spectra of all twelve half-rings are quite similar (see Figure 6).
All show above-solar abundances for most elements (e.g., Si, S, Ca) and
a variation in column density over the range 0.5--$0.9\times10^{21}$
cm$^{-2}$.  The inner three half-rings of the southwest hemisphere
appear brighter than those of the northeast, evidence of which is
apparent from the image as well.  The northeast and southwest halves
of the outer three rings show nearly identical spectra.  The spectra
of the northeast half-rings also show little difference between them,
while the spectra of the southwest half-rings tend to decrease in
brightness moving outward.

The most noticeable differences are in the outermost ring as compared
to all other rings, where the strength of the Fe L-shell blend around
0.73 keV seems diminished.  As compared to the two rings just interior
to it, the continuum in the outermost ring increases by a factor of
$\sim$2.  In terms of emission measure, this is explained by an
increase in the continuum emission (hydrogen thermal bremsstrahlung
in Case H; synchrotron in Case S) rather than simply a decrease in the
emission of Si or Fe.  We verified this by constructing an image of
\e05 in an energy band dominated by continuum emission.  This was
difficult because the continuum dominates the other model components
only in a narrow spectral window around 1.5 keV.  A radial profile of
the continuum image peaks approximately 0.5$\arcsec$ outside of a
similarly constructed profile of the Si K$\alpha$ line.  This confirms
that the continuum emission is more prevalent in the outermost region
of the SNR.

\section{DEPROJECTION}
Again invoking the very circular and distinct shell-like appearance of
\e05, we assume spherical geometry as a good
first approximation to the three-dimensional structure of the remnant.
Using the regions described in the previous section, we did a
deprojection of the remnant to determine the densities of the
different elements for both Case H and Case S.  This is completely
independent of the image fitting procedure described in \S3.2.

Our spectral models fit for the emission measures (EMs) of each
species, which are given by $n_en_iV$, where $n_e$ is the electron
density, $n_i$ is the density of species {\em i}, and $V$ is the
emitting volume.  Our deprojection assumes that these volumes are
portions of spherical shells.  It also assumes that the densities in
each shell are uniform.  In order to determine the quantity $n_en_i$,
we must divide out the emitting volume.  For the outermost ring, this
is straightforward since there is only one shell that contributes to
this ring.  For the innermost ring, all the shells contribute to the
emission.  Therefore, we can populate an upper triangular matrix with
volume elements that, when multiplied by a vector containing the as
yet unknown $n_en_i$ values for the shells, is equal to a vector of
EM$_i$ values for the rings.

Because we defined the rings to have approximately the same number of
events each, rings 4 and 5 were very narrow (about 2 pixels, or
$\sim$1$\arcsec$, thick, see Figure 5).  Since our deprojection
depended on the ``true'' EM in each ring, it was important to take
account of the point-spread-function (PSF) of \chandra, which
additionally is energy dependent.  Using the \chandra\ Ray Tracer
(ChaRT)\footnote{ChaRT and MARX are available at:
http://asc.harvard.edu/chart/index.html}, we created two PSFs: one at
1.85 keV near the Si K$\alpha$ line, and one at 0.75 keV near the Fe
L-shell complex.  For both, we used a ray density of 10 rays/mm$^2$,
which resulted in of order 500,000 photons to define the PSF.  Since
our source is small, we only needed one PSF at each of the two
energies, so we used the center of the remnant as the input location
for ChaRT.  The output of ChaRT was input to the program MARX to
project the rays onto the detector and create images of the PSFs.
Each PSF image was then convolved with an image of an annular ring,
normalized to unity.  This was done for each ring to account for the
amount of flux from one ring that spilled into any other, which was as
much as 18\%.  We must include this in our deprojection since it is
clear the true emission in each ring is ``contaminated'' by emission
from all other rings.  Our matrix now becomes fully populated.

To solve for the densities, we could have done a matrix inversion.
However, this would provide no guarantee that the resulting densities
would be positive definite.  Therefore, we decided instead to fit for
the densities with the requirement that they be $\geq$0.  The
temperature and ionization timescale define the ionization state and
thus the charge state for each species. From this we can calculate the
average number of electrons per ion. With this and the fitted values
for $N_i = n_en_i$, we can obtain the electron density in each shell:
$n_e = (\sum_i N_ic_i)^{1/2}$, where $c_i$ is the charge state for
species $i$.  We can then easily determine the densities, $n_i$, for
each species in each shell.  For both Cases H and S we found that the
densities of the metals were negligible in the inner shells, peaked in
shell 5, and then dropped off again.  In Case H, hydrogen did not
follow this pattern, but instead showed essentially zero density in
shell 5 and then peaked in the outermost shell (see Figure 7).

\section{DISCUSSION}
Our analyses of \e05 point us toward several conclusions which give us
insight into various aspects of SNRs.  We find that the abundances
obtained from our model fits, when compared with theoretical models of
supernovae, confirm that \e05 is the remnant of an SN Ia.  In
addition, the abundance fits are able to give us information as to
which theoretical model is more appropriate.  The global spectrum of
\e05 as well as that of a small knot of material give us important
clues as to the properties of iron in the remnant.  Our deprojection
and image fits provide us with a shell structure to this SNR, which we
use to derive information about the dynamics.  This in turn allows us
to draw the conclusion that the bulk of the continuum emission in this
remnant must be non-thermal in nature.

\subsection{SN Ia Yields}
In Figure 8 we plot our abundances, normalized to Si = 1, for Cases H
(squares) and S (x's).  These yields are also compared to those
determined by two models of SN Ia, the W7 model (fast deflagration,
stars) and the WDD3 model (delayed detonation, circles) of
\citet{iwam}.  The fast deflagration involves a deflagration wave that
propagates outward subsonically, and the delayed detonation scenario
begins as a deflagration that then transitions to a detonation at some
density.  The other models presented by \citet{iwam} are effectively
represented here.  Their W70 model is similar to W7, and the other DD
models are similar to WDD3; therefore we did not plot these.  It is
quite clear that the low-Z elements of O, Ne and Mg are 1--2 orders
of magnitude less abundant than the higher-Z elements of Si - Ca.  This
is a signature of SN Ia's \citep{thiele} and confirms such an origin
for this remnant.

Our yields for O, Ne, and Mg agree better with the WDD3 (delayed
detonation) model than the W7 (fast deflagration) model.  Even if all
of the Si has not been shocked, the low O/Si ratio would only be
exacerbated and the W7 model would still fail to fit the data.
\citet{badenes} have compared various explosion models with the
spectrum of Tycho's SNR.  They are able to show that the pure
detonation and pure deflagration models do not satisfactorily
reproduce important features in Tycho's spectrum.  Their results point
to a delayed detonation explosion as the best model.  In their Figure
4, they plot the integrated EM of each species in the remnant as a
function of time.  In this context, we looked at another Type Ia
remnant, DEM L71.  This SNR is 4400 years old \citep{ghav} and rich in
Si, S, and Fe \citep{hughl71}.  Shocks in older remnants probe
different parts of the SN ejecta.  In comparison to their Figure 4, we
find that at the age of DEM L71 the models which show the same
dominant elements as seen in this SNR are also delayed detonation
models.  This suggests that the delayed detonation scenarios for SN Ia
explosions may be more appropriate.  In addition, all three remnants
indicate some preference for higher transition densities among the
delayed detonation models.  All of these results reveal the power of
remnant studies of Ia's to elucidate explosion mechanism physics.

The iron yields we observe are low compared to the models of
\citet{iwam}.  Our model does underestimate the amount of iron present
in the remnant.  However, it is extremely unlikely that uncertainties
in Fe L-shell atomic physics could account for the difference of a
factor of $\sim$30 between the models and our yield.  We interpret
this discrepancy to be due to the fact that the reverse shock has not
propagated far enough into the remnant to shock all of the iron.  This
is not unprecedented; there is evidence for unshocked iron ejecta in
another young Type Ia, SN1006, where \ion{Fe}{2} absorption lines
have been detected in the center of the remnant \citep{hamilton, wu}.
This is also a feature of theoretical models \citep{badenes}.

One puzzle we encountered was that our fits required iron to be in a
different thermodynamic state than the other elements, with a much
higher temperature and slightly higher ionization timescale.  We
attempted a fit with a component of iron at the same state as the
other elements and found we could allow a maximum abundance of 0.85
for both Cases H and S, which is 0.05 and 0.07, respectively, relative
to Si.  However, such a fit did not reproduce the Fe K line and quite
clearly underestimated the flux from 0.9--1.4 keV.  The fact that
iron appears to be in a different thermodynamic state implies that it
may have come from a different portion of the progenitor than the
other elements.  \citet{hwang} found a similar situation in the case
of Tycho's SNR, where the Fe K emission required a higher temperature,
but a lower ionization timescale than the other elements.
Intuitively, this might be what one would expect if there were a
temperature gradient and the ejecta were stratified, with Fe interior
to the other elements.  We may be seeing an indication of the ejecta
density profile here.  For example, the constant density ejecta
profile of \citet{dc} predicts an increasing temperature gradient
behind the contact discontinuity.  However, it must be emphasized that
X-ray studies probe the electron temperature, while the main thermal
energy rests in the ions.  The extent of the equilibration between
these two components is still under investigation (e.g., Rakowski,
Ghavamian, \& Hughes 2003).  In addition, it is not simple to relate
the values fit for in our models to the {\em actual} temperature and
timescales of the ejecta because of astrophysical complications.  For
example, our fits assume a single, constant temperature for each
species everywhere in the remnant.

\subsection{Iron Enhanced Knot}
We have determined that the knot in the northern interior of \e05 is
enhanced in iron (see Figure 9).  As seen in projection, this knot is
contained mostly in the second northeast half-ring.  We used the
spectrum of that particular ring as a template for the knot's
spectrum.  If we simply renormalize the spectrum, the fit is poor
($\chi^2/d.o.f. = 59.0/32$).  However, if we allow the iron abundance
to be free, we find a good fit at about twice the iron abundance of
the half-ring ($\chi^2/d.o.f. = 32.6/32$).  A fit of pure iron also
gave a poor match to the data, indicating that although this knot is
enhanced in iron, it is not pure iron ejecta.

We looked at this particular knot simply because it was the brightest
and most isolated.  However, given that our image fits (\S 3.2)
show the presence of significant clumpiness in this SNR, it is
possible that other knots show variations in the Si/Fe abundance
ratio, although we are unable to definitively measure this.  Knots
with different X-ray spectra have been detected in Tycho, the
so-called ``typical'' Type Ia SNR, on the southeastern edge of the
remnant \citep{van}.  Recent work has confirmed that the spectral
differences are a result of differences in Si/S and Fe abundances in
these knots \citep{dec}.  In neither Tycho nor \e05 do we see knots of
{\em pure} iron or silicon.  Models of SN Ia explosions (i.e., Iwamoto
et al. 1999) predict that the ejecta are highly stratified, with an
outer O-rich layer, then a Si-rich region, followed by an inner
Fe-rich layer.  However, the boundaries between these regions are not
sharp; the abundances vary smoothly from one zone to the next.  The
differing ratios of Si/Fe observed in the remnants' knots suggest to
us that they originated in the transition region between the iron- and
silicon-rich zones of the ejecta.  Clumping in SN Ia's has been
investigated by \citet{wc}.  They propose that the nickel bubble
effect, in which the radioactive nickel expands and compresses the
shell of material around it, may be responsible for producing knots.
The shell surrounding the nickel would be the Si/Fe transition zone.
Further investigation into possible origins of knots in this region
would be worthwhile.

\subsection{Shell Structure}
Our deprojection analysis clearly indicates the presence of a true
shell structure for this SNR (see Figure 7).  The deprojection was
done for O, Si, S, Fe, and H (the latter only for Case H).  In both
cases, and for both hemispheres, we found that the metals all follow
this structure, with essentially no material in the interior of the
remnant (see below for a discussion of the southwest hemisphere).  The
metals extend over the two outermost shells in the deprojection with a
peak density in the shell at $R \sim 3.2$ pc and a slightly lower
density in the outer shell at $R \sim 3.6$ pc.  Hydrogen (Case H)
shows a slightly different structure: there is almost no H in the $R
\sim 3.2$ pc shell or further in the interior; rather the continuum
comes from the outermost shell.  The continuum emission in Case S,
when deprojected, also comes only from the outermost shell.  In each
case, therefore, this is clear evidence for a blast-wave component to
the remnant.  We note however that the blast wave component appears to
be contaminated somewhat by ejecta, presumably by fragments that
preceed the main shell.

In the southwest hemisphere, the structure is more complicated.  There
is a shell at a radius of $\sim$ 3.2 pc that is the counterpart to the
northeast shell.  In addition, there appears to be another spectral
component peaking at a smaller radius, with an apparent gap between
the two shells.  We do not believe this is an actual shell, but rather
a projection effect due to the clumpy region on the southwest side.
Given simply a shell of material, one would expect to see a radial
profile like that of the northeast hemisphere.  Now if one adds a
clump interior to the shell, the radial profile would include a
``bump'' at some inner radius, as seen in the southwest hemisphere.
The radius at which this bump appears gives no indication of the true
position of the clump, which could be almost anywhere along the
line-of-sight in the remnant.  Our simple deprojection analysis
ignores this fact, and assumes that the emission from the clump is
spread over an inner shell.  Therefore, the density derived for these
inner portions is clearly an overestimate for the interior of the
remnant.  On the other hand, it may be an underestimate for the true
density of the clump itself.  We believe that the clumpy region in the
southwest of \e05 is a result of enhanced density in or a deeper
penetration of the reverse shock into a portion of the ejecta
shell. This could be caused by a small region of enhanced ambient
density or by intrinsic asymmetry in the explosion process itself.

Because our deprojection analysis revealed the shell-like structure of
this remnant, we can be confident that the image fits, which assumed
such a structure, produced meaningful results.  The fit results give
us a constraint on the evolutionary state of the remnant if we
identify the outer edge of the X-ray emitting shell with the location
of the blast wave, and the inner edge with the location of the reverse
shock.  In the context of self-similar models for the evolution of
young SNRs (e.g., Truelove \& McKee 1999), the observed ratio of these
radii (1.185--1.192, including errors), corresponds to a particular
point (or range of points) on the scaled radius vs.\ time curve. These
curves depend on the assumed radial density distribution of the
ejecta; we consider the cases of constant and exponential profiles.
From the constraint on the scaled radius vs.\ time curve and the
physical radius of the remnant, we obtain a relation between the
ejected mass and the density of the ambient medium.  Assuming a
Chandrasekhar mass for the ejecta, the density is
$\sim$4--$5\times10^{-2}$ cm$^{-3}$ for the constant density ejecta
profile of \citet{tm}, and $\sim$0.7--$1\times10^{-2}$ cm$^{-3}$ for
the exponential density ejecta profile of \citet{dc}. These estimates
imply a low density environment and yield swept-up masses for hydrogen
of 0.055--$0.068~{\rm M{_\sun}}$ or 0.0095--$0.013~{\rm M{_\sun}}$,
respectively.  The evolutionary models also can provide estimates of
the explosion energy and age of the remnant, but they rely on knowing
the shock velocity at the current epoch, which is only constrained to
be $>$3600 km s$^{-1}$. The explosion energy is $>$$0.15\times
10^{51}\,(V_s/3600\,\rm km\, s^{-1})^2$ erg (constant density) and
$>$$0.09\times 10^{51}\,(V_s/3600\,\rm km\, s^{-1})^2$ erg
(exponential). The age is $<$$860\,(V_s/3600\,\rm km\, s^{-1})^{-1}$
yr (constant density) and $<$$670\,(V_s/3600\,\rm km\, s^{-1})^{-1}$
yr (exponential).

\subsection{Non-Thermal Continuum}
\e05 is believed to be situated in a low density environment.  There
is the dynamical argument just given, which constrains the density to
$n_{\rm H} \lesssim0.05$ cm$^{-3}$.  \citet{tuohy} obtained a value of
$n_{\rm H} \lesssim0.02$ cm$^{-3}$ through optical studies.  In
addition, the remarkable symmetry of the outer shell of this SNR
suggests the ambient density is much lower than that surrounding other
remnants (given the same level of density fluctuations).  For example,
we do not see breaks in the rim or an irregular shape in \e05 as we do
in Tycho \citep{hwang02}, which has an ambient density of $n_{\rm H}
\sim0.5$ cm$^{-3}$ \citep{hughes}.  Our Case H deprojection yields a
post-shock density of $3.6~{\rm cm}^{-3}$ for hydrogen, which in turn
yields a swept-up hydrogen mass of 7.4 ${\rm M{_\sun}}$.  This
density, converted to its preshock value of $\sim$1 cm$^{-3}$ assuming
the strong shock compression factor of 4, is much too high to be
consistent with the previous arguments. If we take the ambient density
to be $n_{\rm H}=0.05$ cm$^{-3}$, then the level of hydrogen thermal
continuum produced would be approximately $3\times 10^{-3}$ times
lower than the {\em actual} continuum level observed in the
spectrum. Therefore we are led to conclude that there is a significant
non-thermal component of continuum emission in this SNR.

In Table 4 we show the masses of H, O, Si, S, and Fe in the two
outermost shells, derived from the densities obtained by our
deprojection analysis under spherical geometry.  Although the masses
derived from the non-thermal case are too high to be consistent with
SNe Ia (0.76 M$_\sun$ for Si), this is a limiting case since it
assumes {\em all} the electrons come from the partially ionized
metals.  Therefore, it is an extreme upper estimate to the mass.  If
even a small amount of hydrogen were included with the non-thermal
emission, the densities, and hence masses, of the metals would
decrease.  In addition, the clumpy nature of the SNR inferred from our
image fits (\S 3.2) suggests that the apparently diffuse emission
(from the shell component in the image fits) may arise from a smaller
volume than expected if it were uniformly spread throughout.  The
knots modeled in the image fits produce $\sim$20\% of the X-ray events
from the shell in the northeast half of the remnant, while occupying
only $\sim$2\% of the shell volume of the diffuse emission.  If we
assume that the diffuse component is made up of similar, though
unresolved, knots with the same X-ray events-to-volume ratio as those
fit directly, we find that only $\sim$10\% of the diffuse emission
volume need be occupied by knots to produce the same emission.  This
10\% filling factor translates to a $\sim$30\% reduction in the
derived masses, bringing the values for the non-thermal case to a more
reasonable level: about 0.2 M$_\sun$ for Si.

Several other young, shell-like SNRs show evidence for non-thermal
emission \citep{koy2,allen,slane}, the most well-known being SN1006
\citep{koy}.  The shell of SN1006 is dominated by non-thermal
emission, showing a featureless spectrum that can be described by a
power-law with a photon index of 2.95 \citep{koy}.  The synchrotron
cut-off model was also applied to SN1006 and a value of
$6\times10^{16}$ Hz for the cut-off frequency was obtained
\citep{reykeo}.  For \e05 we find a higher power-law photon index
($\alpha_p\sim3.3$) and a lower synchrotron cut-off frequency
($1.6\times10^{16}$ Hz).  Both of these results are consistent with
\e05 showing a steeper non-thermal spectrum in the 0.5--7 keV X-ray
band than SN1006.  As to the overall intensity of non-thermal
emission, these two remnants are quite similar as well, with \e05
being about a factor of three more luminous than SN1006.  Other SNRs,
such as Tycho, Kepler, and Cas A, are not dominated by non-thermal
emission but are well-fit by a power-law at energies $\gtrsim$10 keV
(Allen et al. 1999).  In the case of \e05, we get as good, if not a better,
fit for the global spectrum with a non-thermal continuum.  Thus a
non-thermal origin for the continuum emission from \e05 is plausible.

We also applied the minimum-energy condition for the synchrotron
emission (Longair 1994, pp.~292-296) from this remnant, assuming equal
energy densities for the protons and electrons and a volume filling
factor of one.  By plotting the extrapolated radio power-law against
the cut-off power-law determined by the synchrotron cut-off model, we
found that the latter begins to deviate from a straight power-law at
$\sim$10$^{14}$ Hz. We used this value as the maximum frequency in our
calculation and note that varying this value did not change our
results greatly. We obtained a minimum energy of $10^{48}$ ergs and
magnetic field of 60 $\mu$G.  The energy requirements on this
component are safely below the available energy of $10^{51}$ ergs, but
our magnetic field value is large.  However, the minimum energy
condition (effectively equipartition between particles and the
magnetic field) is unlikely to be applicable, as shown by Dyer et
al.~(2001) for SN1006 from comparing TeV gamma-ray and non-thermal
X-ray emission.  If we assume a magnetic field of $\sim$ 10 $\mu$G as
they find, then the total energy required to explain the non-thermal
emission in 0509-67.5 increases to $10^{49}$ ergs, still well below
the available energy.  Using $\sim$ 10 $\mu$G for the magnetic field
along with the fitted value of the roll-off frequency from the
synchrotron cut-off spectral model, we find the maximum energy of
electrons to be $\sim$ 20 TeV.

The preceding arguments strongly support the picture that the forward
shock of \e05 is accelerating particles to relativistic energies.
\e05 is thus the first Magellanic Cloud SNR for which such an energetic
cosmic ray component has been securely detected.  One task that
remains to be done to make our work fully consistent is to explore the
effects of modifications to the remnant dynamics due to efficient
particle acceleration. As the fraction of the shock front's kinetic
energy being diverted to the acceleration of relativistic particles
grows, the thicknesses of the forward and reverse shocked regions tend
to decrease \citep{dec2}. Such changes could affect our estimate of
the ambient medium density based on the thickness of the X-ray
emitting shell.  In the absence of a specific model for \e05, it is
not possible to estimate a correction factor for the ambient density.
Although beyond the scope of this work, development of such a model
would be a valuable next step toward determining the efficiency of
cosmic ray acceleration in \e05.

\section{SUMMARY}
\e05 has provided tantalizing clues to the physics of SNe and SNRs.
We found that the remnant abundances are consistent with those
produced by a Type Ia SN, with a low O/Si ratio.  The best-fit relative
abundances of O, Ne, Mg, Si, S, Ar and Ca, when compared with models
of SN Ia explosions, agree best with the yields from delayed
detonation models for the propagation of the burning front in the
SN. Fast deflagration models, such as the well known W7 model, are in
poorer agreement with our data.  The X-ray iron emission from this
remnant poses something of a puzzle.  Its abundance relative to Si
falls far below what we expect from SN Ia models, and from this we
infer that the ejecta are compositionally stratified and the reverse
shock has not yet propagated deeply into the Fe-rich zone of the
remnant.  There is potentially more unshocked, cold iron in the center
of the SNR.  The iron emission also prefers an apparently different
thermodynamic state than the other elements, requiring a higher
temperature for an acceptable fit to the spectrum.  This has also been
seen in Tycho's SNR, and may be an indication that the shocked iron
comes from a different portion of the progenitor than the other
elements. This would additionally require that there be a radial
temperature gradient in the reverse-shock-heated ejecta.  An
isolated knot in the SNR was found to be relatively enhanced in iron
compared to its surrounding region.  Along with similar knots found in
Tycho's SNR, which also show variations in the Si/Fe ratio, this may
point to an origin for the knots in the transition zone between the
iron- and silicon-rich ejecta.

A deprojection analysis to determine the three-dimensional structure
of \e05 indicates that the metals are largely contained within a
spherical shell that peaks interior to the deprojected continuum
emission.  Specifically, the ejecta extend over the outer two shells,
while the continuum comes only from the outermost one.  From this we
conclude that the continuum emission from \e05 comes from the blast
wave and that the blast wave region is contaminated by some ejecta
material.  If we assume that the continuum comes from thermal
bremsstrahlung emission of hydrogen, we obtain a preshock density for
the ambient medium ($n_{\rm H} \sim 1$ cm$^{-3}$) that is a factor of
20 times greater than the density ($n_{\rm H} \sim 0.05$ cm$^{-3}$)
derived from the remnant's dynamical state and other arguments. On the
other hand, a model where the continuum emission arises from
non-thermal synchrotron radiation from shock-accelerated relativistic
electrons provides as good if not a better fit to the
spectrum. Furthermore, the luminosity, spectral shape, and overall
energy requirements of the synchrotron emission are consistent with
that of other young SNRs, like SN1006.  We conclude therefore that (1)
the ambient density surrounding \e05 is low and the contribution of
swept up matter to the observed spectrum is negligible, and (2) the
continuum emission is virtually all non-thermal and bears a close
similarity to the non-thermal emission from SN1006.  This makes \e05
another member of the class of SNRs showing evidence for a significant
population of relativistic electrons. It is somewhat remarkable to
find evidence for the need of a non-thermal component in the 0.2--7
keV \chandra\ band, particularly for a spectrum that is so clearly
dominated by emission lines.  This provides another avenue for
investigating the origins of non-thermal emission and the processes,
e.g., diffusive shock acceleration, that give rise to it in SNRs.

\acknowledgements

We would like to thank Pat Slane for numerous science discussions and
comments on the manuscript, Paul Plucinsky for help with obtaining the
data on the calibration sources, John Nousek and Dave Burrows for
assistance with the original proposal, R.~Chris Smith for providing
the H$\alpha$ image of \e05, Vikram Dwarkadas for numerical simulation
results and the anonymous referee for helpful comments. This research
was partially supported by \chandra\ grants GO2-3069X, GO2-3070X, and
GO3-4086X.

\clearpage

\begin{deluxetable}{cccccc}
\tablewidth{0pt}
\tablecolumns{6}
\tablecaption{Calibration Line Centroids}
\tablehead{\colhead{Line} & \colhead{14-15 May} & \colhead{14 May} & \colhead{12 May} & \colhead{11-12 May} & \colhead{Accepted}\\ \colhead{~} & \colhead{62055} & \colhead{62056} & \colhead{62057} & \colhead{62058}}
\startdata
Al K$\alpha$ & $1.489^{+0.002}_{-0.002}$ & $1.484^{+0.003}_{-0.003}$ & $1.489^{+0.003}_{-0.003}$ & $1.489^{+0.005}_{-0.004}$ & 1.48656\\
Ti K$\alpha$ & $4.502^{+0.016}_{-0.001}$ & $4.501^{+0.007}_{-0.001}$ & $4.502^{+0.006}_{-0.002}$ & $4.502^{+0.036}_{-0.001}$ & 4.50885\\
Mn K$\alpha$ & $5.899^{+0.001}_{-0.001}$ & $5.895^{+0.005}_{-0.006}$ & $5.895^{+0.005}_{-0.003}$ & $5.899^{+0.001}_{-0.003}$ & 5.89505\\
Mn K$\beta$ & $6.497^{+0.008}_{-0.008}$ & $6.499^{+0.003}_{-0.012}$ & $6.497^{+0.005}_{-0.010}$ & $6.502^{+0.014}_{-0.003}$ & 6.49045\\
\enddata
\tablecomments{Energies in keV; errors are 1$\sigma$.  Accepted values from \citet{crc}.}
\end{deluxetable}

\begin{deluxetable}{ccc}
\tablewidth{0pt}
\tablecaption{Best Fit Elemental Abundances}
\tablecolumns{3}
\tablehead{\colhead{~} & \colhead{High energy band}  &
\colhead{Entire band} \\
\colhead{Species} & 
\colhead{($1.6\, {\rm keV} < E < 7.5\, {\rm keV}$)}  &
\colhead{($0.2\, {\rm keV} < E < 7.5\, {\rm keV}$)}}
\startdata
O  & \nodata                & $0.25\pm0.04$ \\
Ne & \nodata                & $<$ 0.05      \\
Mg & \nodata                & $0.25\pm0.07$ \\
Si & $18.5\pm2.6$           & $16.5\pm2.1$  \\
S  & $22.4\pm3.9$           & $15.5\pm2.9$  \\
Ar & $15.3\pm9.2$           & $7.0\pm4.4$   \\
Ca & $32.7^{+35.4}_{-24.6}$ & $5.9\pm2.1$   \\
Fe & $0.81\pm0.16$          & $1.11\pm0.20$ \\
\enddata
\tablecomments{Abundances with respect to solar; quoted errors 
at 90\% confidence level.}
\end{deluxetable}

\begin{deluxetable}{ccc}
\tablewidth{0pt}
\tablecaption{Best-Fit Global Spectral Parameters}
\tablecolumns{3}
\tablehead{\colhead{~} & \colhead{Thermal} & \colhead{Non-thermal}\\ 
\colhead{Parameter} & \colhead{continuum (H)} & \colhead{continuum (S)}}
\startdata
$N_{\rm H}$ ($10^{21}$ cm$^{-2}$) & $0.70\pm0.06$ & $0.61\pm0.06$ \\
$kT$ (keV) & $2.23\pm0.29$ & $3.13\pm0.55$ \\
$\log(n_et/{\rm cm^{-3} s})$ - Si & $9.94\pm0.02$ & $9.93\pm0.02$ \\
$\log(n_et/{\rm cm^{-3} s})$ - others & $10.19\pm0.05$ & $10.14\pm0.04$ \\
$kT$ (keV) - Fe & $10.00_{-4.14}$ & $10.00_{-5.44}$ \\
$\log(n_et/{\rm cm^{-3} s})$ - Fe & $10.53\pm0.02$ & $10.53\pm0.02$ \\
$n_en_{\rm H}V$ ($10^{58}$ cm$^{-3}$) & $2.87\pm0.50$ & 0. \\
$n_en_{\rm O}V/$[O/H]$_\sun$ ($10^{58}$ cm$^{-3}$)   & $0.72\pm0.05$ & $0.63\pm0.06$ \\
$n_en_{\rm Ne}V/$[Ne/H]$_\sun$ ($10^{58}$ cm$^{-3}$) & $<$ 0.15      & $<$ 0.09 \\
$n_en_{\rm Mg}V/$[Mg/H]$_\sun$ ($10^{58}$ cm$^{-3}$) & $0.72\pm0.14$ & $0.43\pm0.14$ \\
$n_en_{\rm Si}V/$[Si/H]$_\sun$ ($10^{58}$ cm$^{-3}$) & $47.4\pm6.7$  & $34.7\pm5.2$ \\
$n_en_{\rm S}V/$ [S/H]$_\sun$($10^{58}$ cm$^{-3}$)   & $44.5\pm6.7$  & $36.5\pm5.2$ \\
$n_en_{\rm Ar}V/$[Ar/H]$_\sun$ ($10^{58}$ cm$^{-3}$) & $20.1\pm12.1$ & $25.9\pm8.9$ \\
$n_en_{\rm Ca}V/$[Ca/H]$_\sun$ ($10^{58}$ cm$^{-3}$) & $17.0\pm5.3$  & $16.7\pm4.3$ \\
$n_en_{\rm Fe}V/$[Fe/H]$_\sun$ ($10^{58}$ cm$^{-3}$) & $3.19\pm0.07$ & $2.96\pm0.09$ \\
$\nu_{cut-off}$ ($10^{16}$ Hz)  & \nodata & $1.60\pm0.10$ \\
$\chi^2/d.o.f.$ & 167.9/167 & 140.2/167 \\
\enddata
\tablecomments{$\alpha=0.46$, SRCUT norm=0.066 Jy, in accordance with radio data; quoted errors are for a 90\% confidence level, and $\chi^2/d.o.f.$ includes the systematic error.}
\end{deluxetable}

\begin{deluxetable}{ccc}
\tablewidth{0pt}
\tablecaption{Masses of Elements}
\tablecolumns{3}
\tablehead{\colhead{Element} & \colhead{Thermal} & \colhead{Non-thermal}\\ \colhead{~} & \colhead{continuum (H)} & \colhead{continuum (S)}}
\startdata
H  & $7.4$  & \nodata \\
O  & $0.08$ & $0.17$ \\
Si & $0.46$ & $0.76$ \\
S  & $0.16$ & $0.37$ \\
Fe & $0.07$ & $0.12$ \\
\enddata
\tablecomments{Masses of outer two shells in units of M$_\sun$, based on best-fit density values and assuming spherical shells.}
\end{deluxetable}

\clearpage

\begin{figure}
\plotone{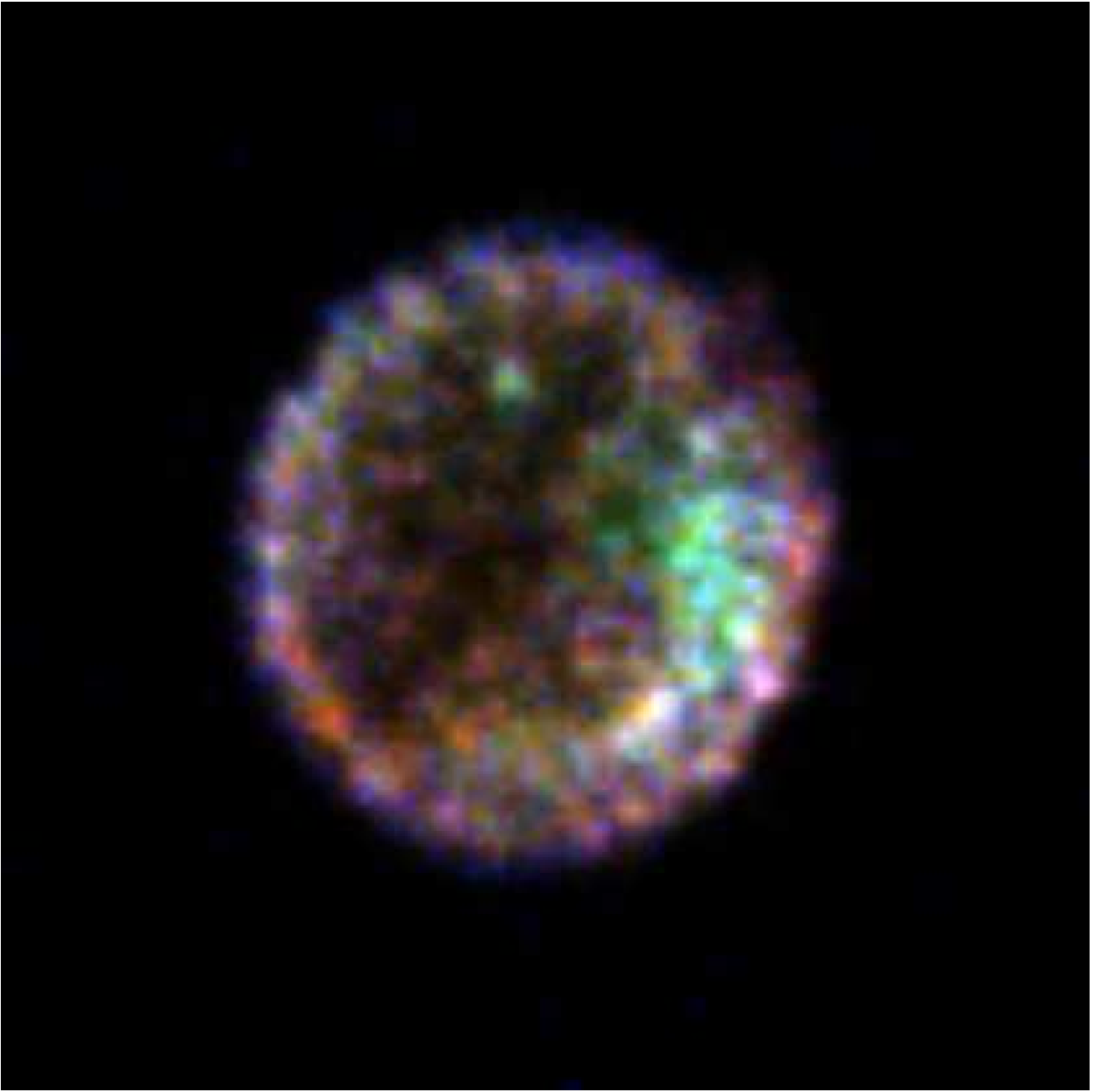} 

\caption{The 3-color \chandra\ broadband (0.2--7 keV) X-ray image of
\e05.  The image has been smoothed with a gaussian with a sigma of
$\sim0.25\arcsec$.  Red contains emission from 0.2--0.69 keV, green
0.69--1.5 keV, and blue 1.5--7 keV.  North is up and east is to the
left.}
\end{figure}

\begin{figure}
\plotone{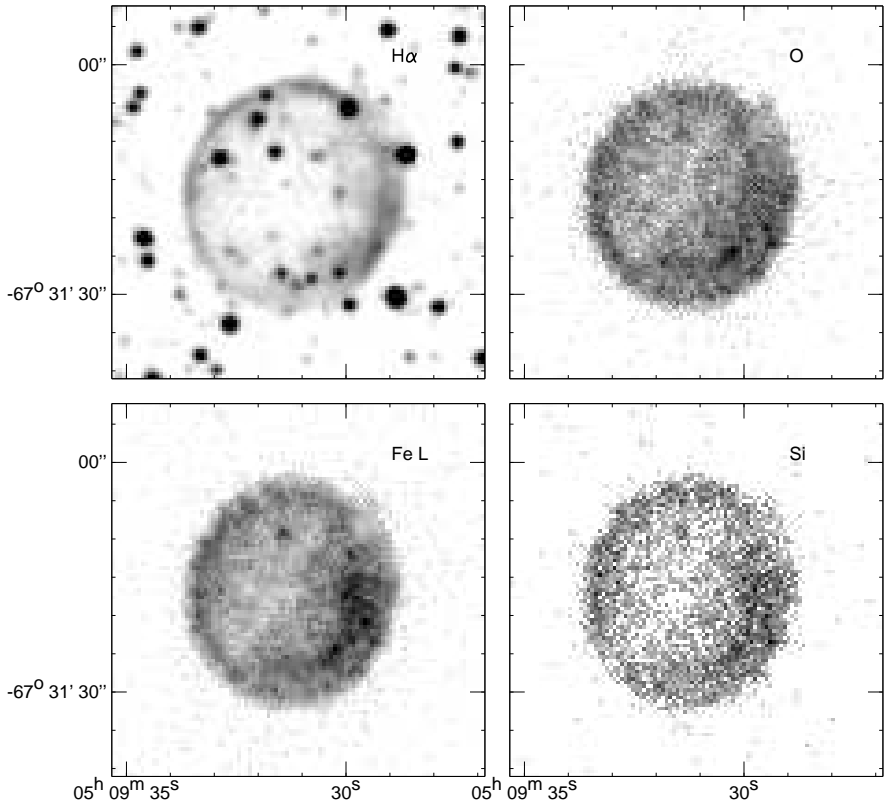} 

\caption{CTIO 4-m H$\alpha$ image of \e05 (\textit{upper left}),
\chandra\ Oxygen 0.45--0.7 keV image (\textit{upper
right}), Iron L 0.7--1.4 keV image (\textit{lower left}), and Silicon
1.5--2 keV image (\textit{lower right}).  The intensity to grayscale 
mapping is square root.  North is up and east is to the left.}
\end{figure}

\begin{figure}
\plotone{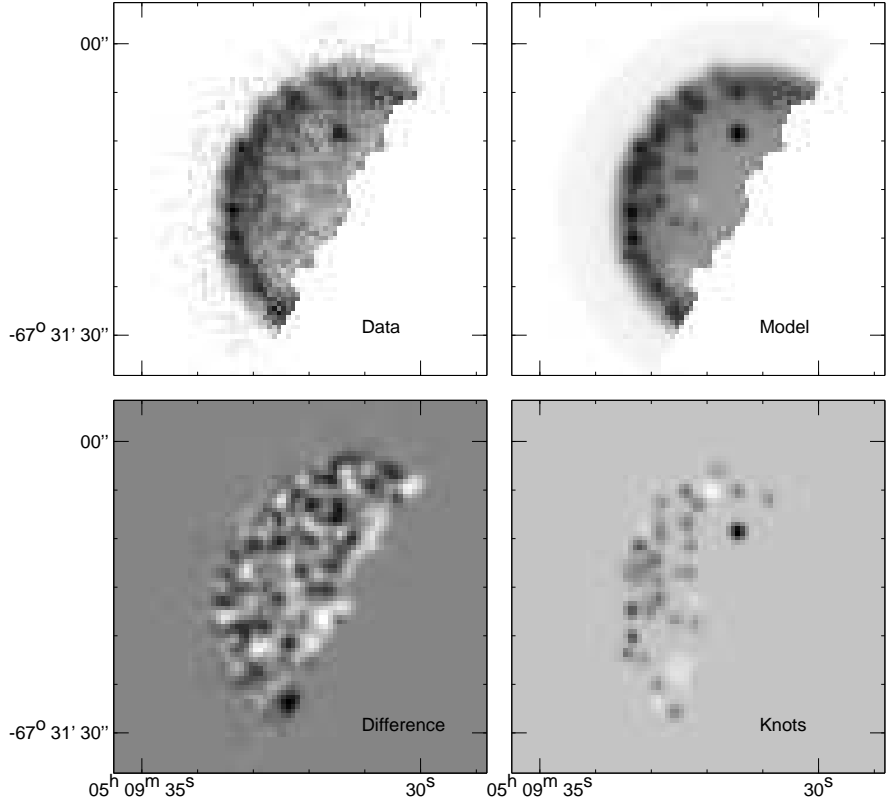} 

\caption{Image fits to the northeast half of the remnant.  The
upper left panel shows the broadband \chandra\ image.  The model,
including knots, is shown in the upper right, and the difference
between the model and the data is in the lower left.  The lower right
panel displays the knot model alone. The intensity range of the grayscale
in the data and model panels is 0 to 100 X-ray events per pixel, the
difference panel is $-$10 to 10 events per pixel, while the knots
panel ranges over values of $-$10 to 90 events per pixel. North is up
and east is to the left.}
\end{figure}

\begin{figure}
\plotone{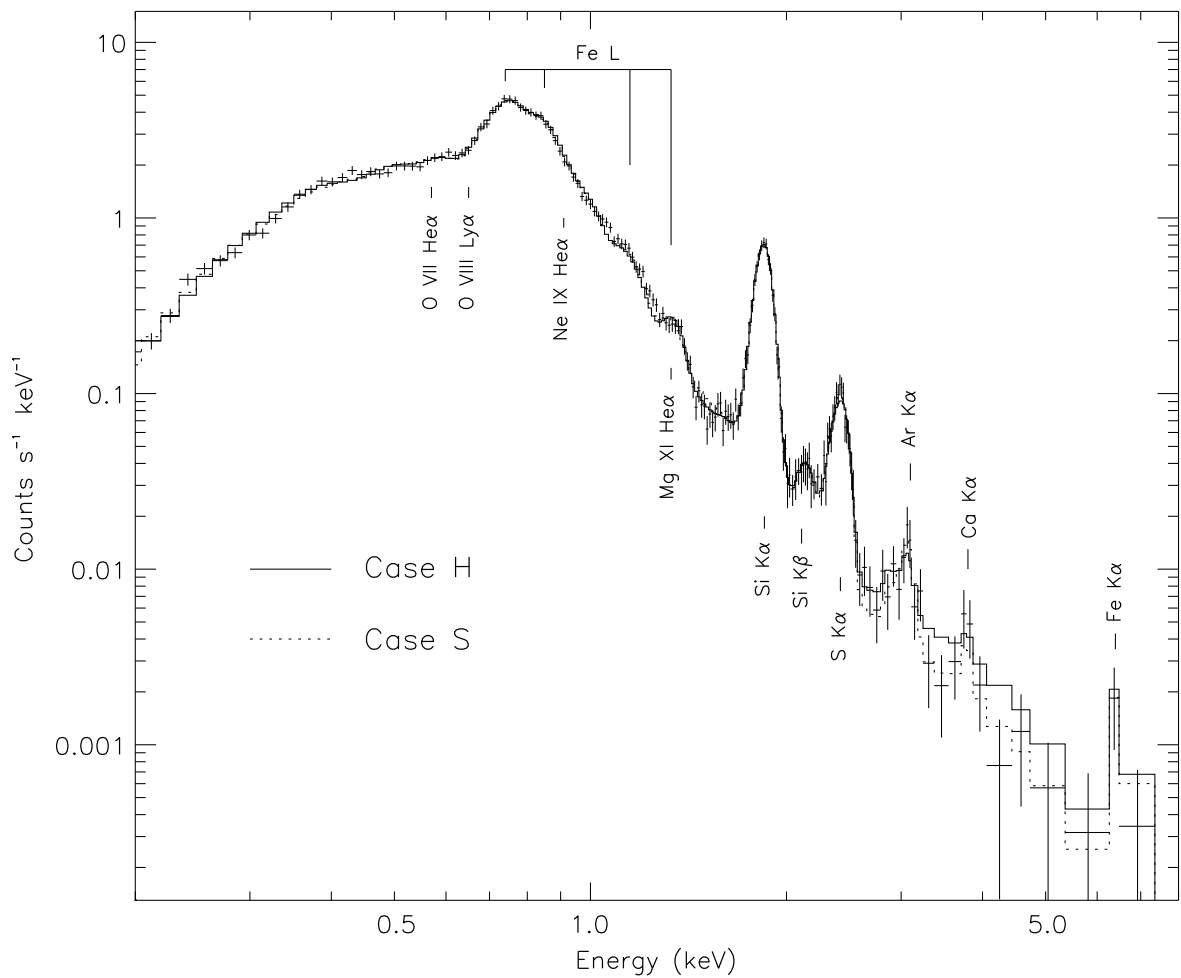} 

\caption{Integrated spectrum of \e05.  Crosses are actual data.
Prominent spectral lines are labeled; the positions of lines from O,
Ne, and Mg are also labeled. Although these lines are not clearly
resolved in the data, our spectral fits require them.  Solid curve is
the hydrogen continuum model (thermal - Case H), and dotted curve is
the synchrotron cut-off (non-thermal - Case S) model.  Note that the
difference in the two models only becomes apparent at the highest
energies, where the synchrotron model seems to fit slightly better.}
\end{figure}

\begin{figure}
\plotone{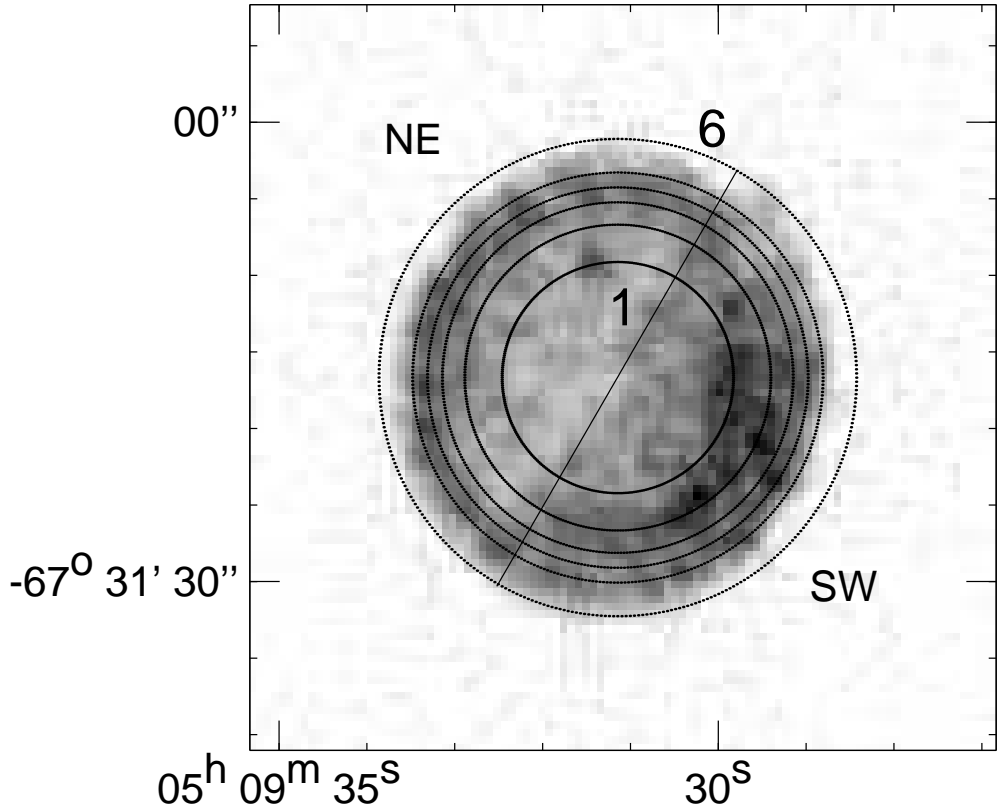}

\caption{Broadband 0.2--7 keV \chandra\ X-ray image of \e05
overlaid with the ring regions used.  The rings are labeled from
innermost (1) to outermost (6).  The two hemispheres are denoted by
northeast (NE) and southwest (SW).  North is up and east is to the
left.}
\end{figure}

\begin{figure}
\plotone{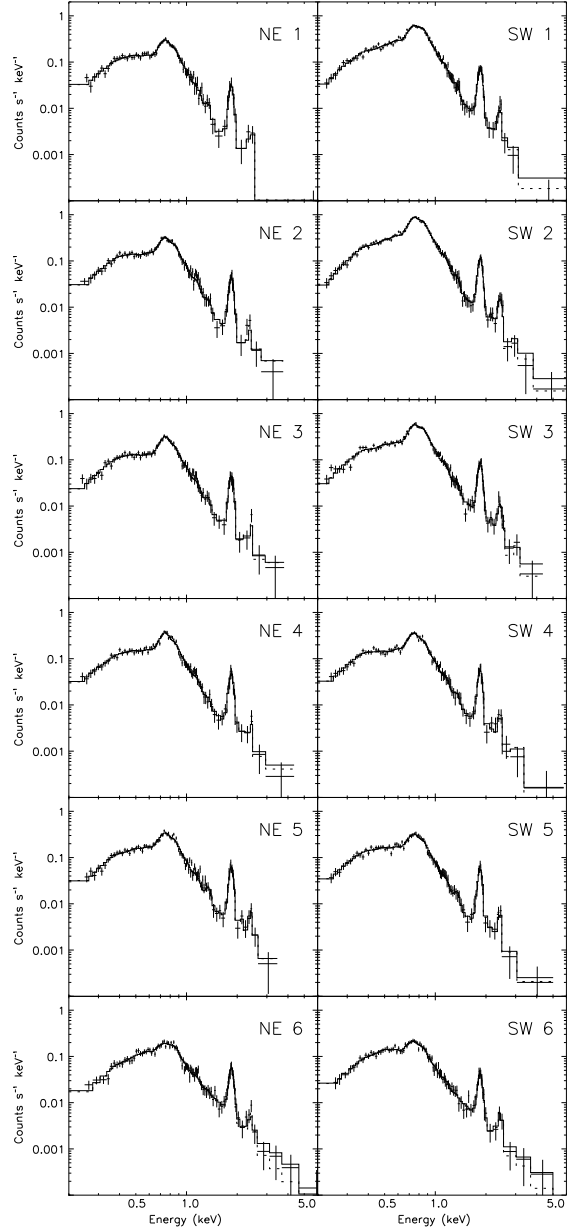}

\caption{Spectra of the 12 half-rings.  NE refers to the northeast
hemisphere; SW refers to the southwest hemisphere.  The solid curve is
the thermal continuum (Case H) model, and the dotted curve is the
non-thermal continuum (Case S) model.  The rings are labeled from
innermost (1) to outermost (6).}
\end{figure}

\begin{figure}
\epsscale{0.9}
\plotone{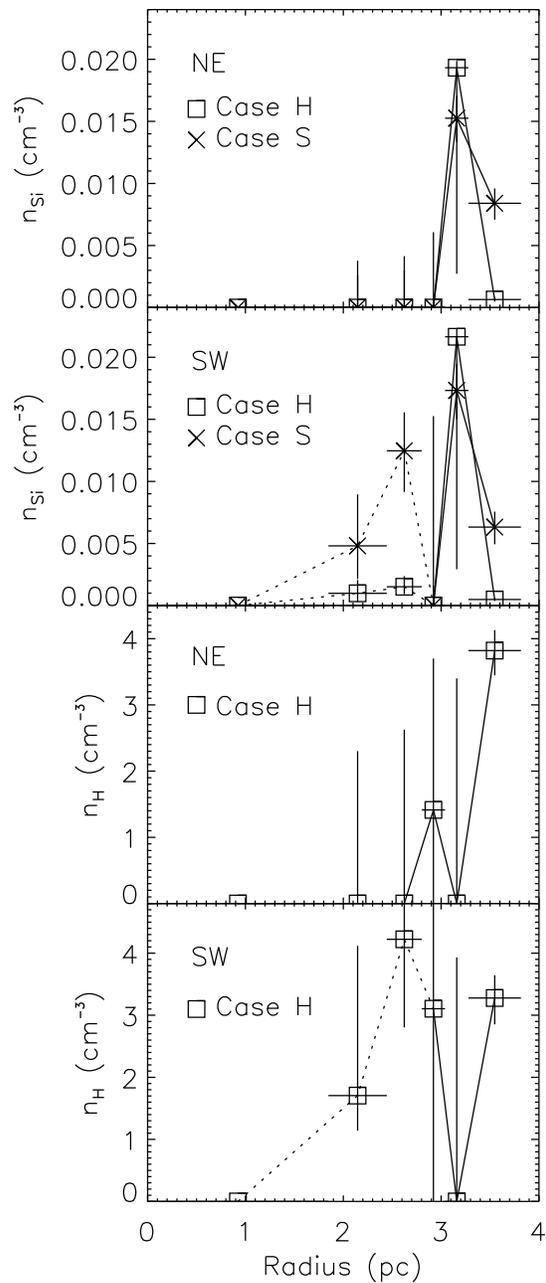} 

\caption{\footnotesize Densities of silicon (\textit{top 2 panels}) and hydrogen
(\textit{bottom 2 panels}) as a function of radius for both the
northeast and southwest hemispheres (as defined in \S 4.2).  The
dotted line is not a second shell, but rather a projection effect due
to the clumpy region of emission seen on the southwest side of
the image.  The innermost shell was not fit.}
\end{figure}

\begin{figure}
\epsscale{1}
\plotone{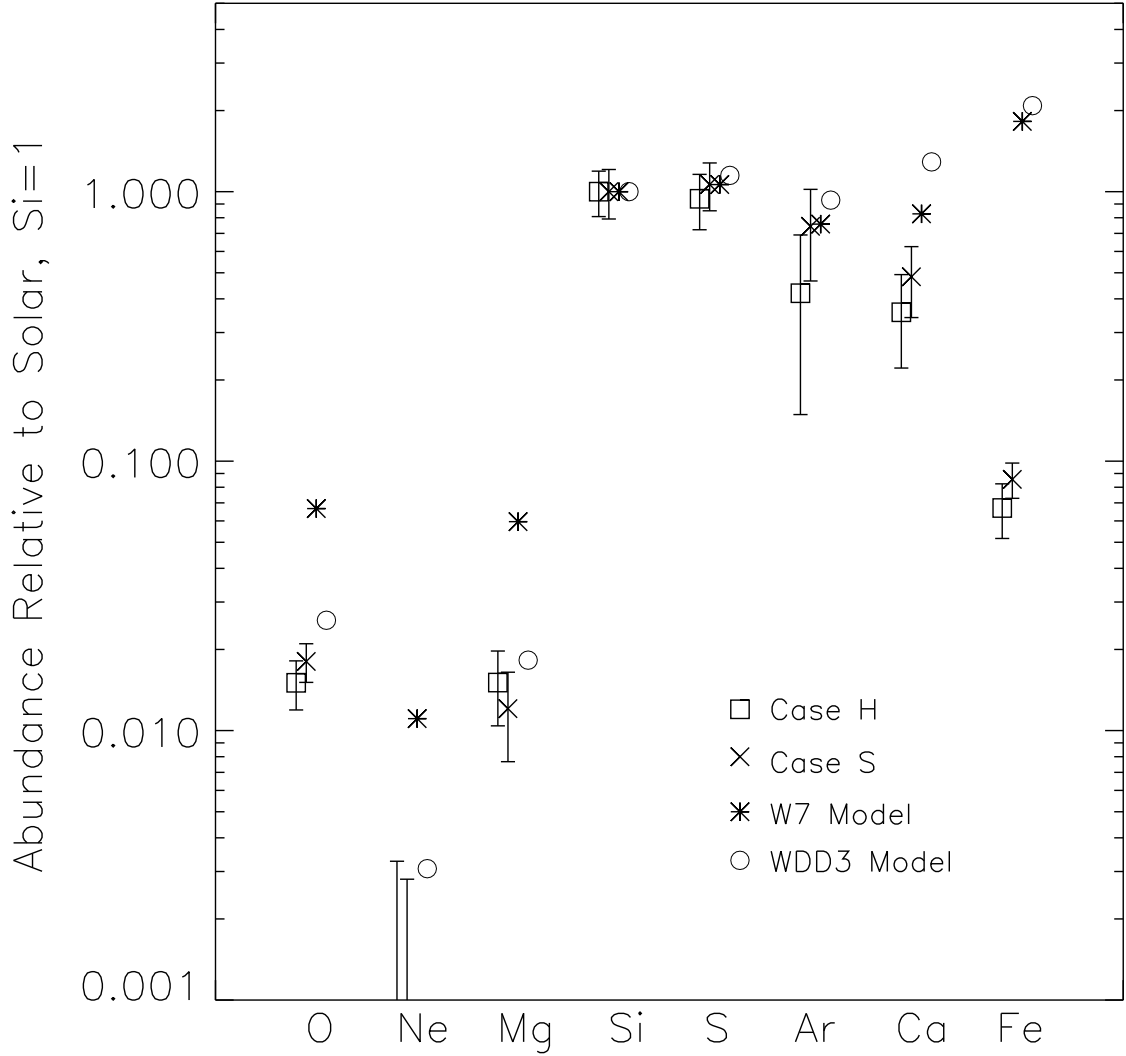} 

\caption{Abundances of the elements, relative to solar and normalized
to Si = 1, from the full 0.2--7 keV band.  Errors are 1$\sigma$.  The
W7 and WDD3 models are from \citet{iwam} and are each normalized to Si
= 1.}
\end{figure}

\begin{figure}
\plotone{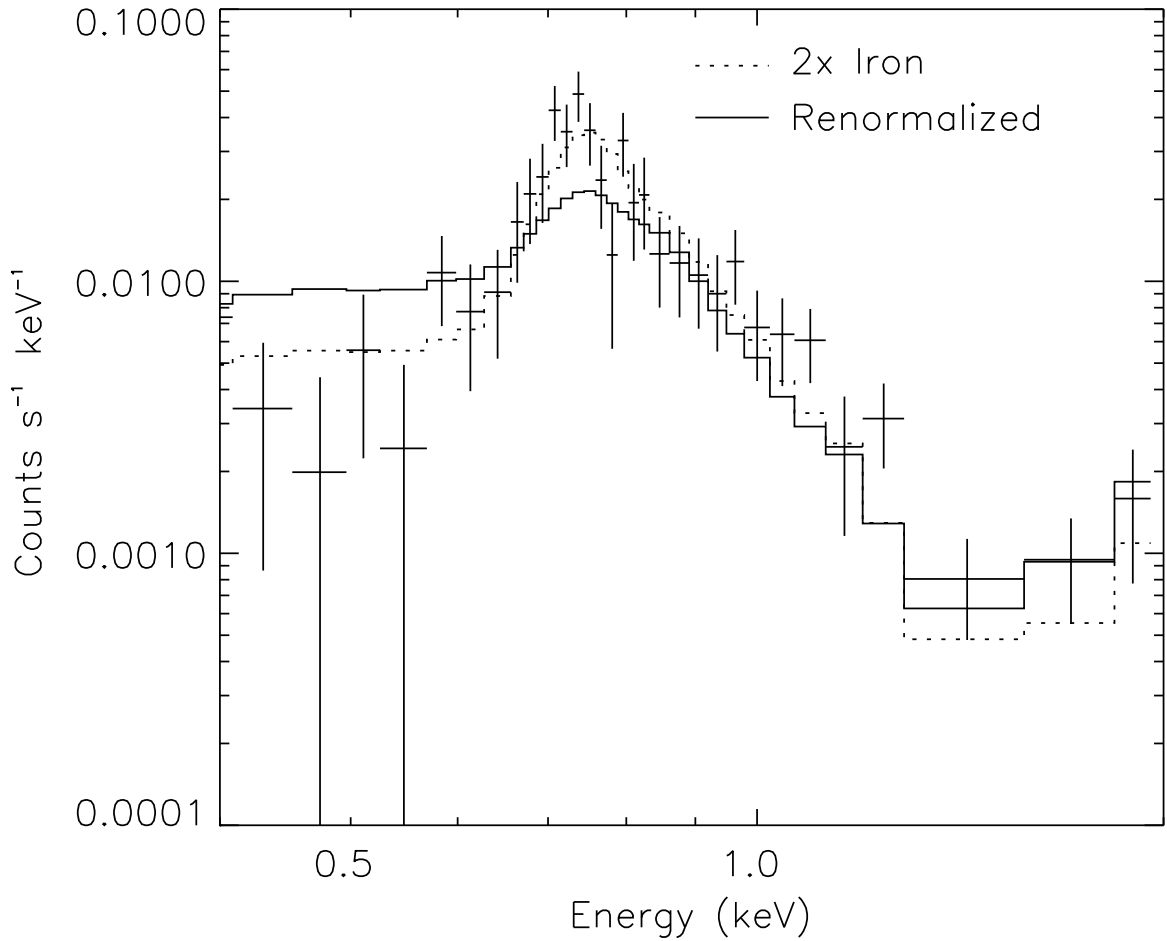} 

\caption{Spectrum of knot in north of remnant.  Crosses are actual
data.  Solid line is the renormalized model for the ring which
contains the knot, and dotted line is the same but with twice the iron
abundance.  The latter model fits much better, indicating that this is
an iron-enriched knot of material.}
\end{figure}


\begin{thebibliography}{}

\bibitem[Allen, Gotthelf, \& Petre(1999)]{allen} Allen, G. E., Gotthelf, E. V., \& Petre, R. 1999, in Proc. 26th Internatl. Cosmic Ray Conf., ed. D. Kieda, M. Salamon, \& B. Dingus, 3, 480 (http://krusty.physics.utah.edu/$\sim$icrc1999/root/icrc.html)

\bibitem[Allen et al.(1997)]{allen97} Allen, G. E. et al. 1997, ApJ, 487, L97

\bibitem[Anders \& Grevesse(1989)]{ag} Anders, E., \& Grevesse, N. 1989, GeCoA, 53, 197

\bibitem[Badenes et al.(2003)]{badenes} Badenes, C., Bravo, E., Borkowski, K. J., \& Dom\'{i}nguez, I. 2003, ApJ, 593, 358

\bibitem[Branch et al.(1995)]{branch} Branch, D., Livio, M., Yungelson, L. R., Boffi, F. R., \& Baron, E. 1995, PASP, 107, 1019

\bibitem[Decourchelle et al.(2001)]{dec} Decourchelle, A. et al. 2001, A\&A, 365, L218

\bibitem[Decourchelle, Ellison, \& Ballet(2000)]{dec2} Decourchelle, A., Ellison, D. C., \& Ballet, J. 2000, ApJ, 543, L57

\bibitem[Dickey \& Lockman(1990)]{dl} Dickey, J. M. \& Lockman, F. J. 1990, ARA\&A, 28, 215

\bibitem[Dwarkadas \& Chevalier(1998)]{dc} Dwarkadas, V. V. \& Chevalier, R. A. 1998, ApJ, 497, 807

\bibitem[Dyer et al.(2001)]{dyer} Dyer, K. K., Reynolds, S. P., Borkowski, K. J., Allen, G. E., \& Petre, R. 2001, ApJ, 551, 439

\bibitem[Ghavamian et al.(2003)]{ghav} Ghavamian, P., Rakowski, C. E., Hughes, J. P., \& Williams, T. B. 2003, ApJ, 590, 833

\bibitem[Hamilton \& Fesen(1988)]{hamilton} Hamilton, A. J. S. \& Fesen, R. A. 1988, ApJ, 327, 178

\bibitem[Hendrick \& Reynolds(2001)]{henrey} Hendrick, S. P. \& Reynolds, S. P. 2001, ApJ, 559, 903

\bibitem[Hillebrandt \& Niemeyer(2000)]{hillnie} Hillebrandt, W. \& Niemeyer, J. C. 2000, ARA\&A, 38, 191

\bibitem[Hughes et al.(1995)]{hughes95} Hughes, J. P. et al. 1995, ApJ, 444, L81

\bibitem[Hughes \& Birkinshaw(1998)]{hughbirk} Hughes, J. P. \& Birkinshaw, M. 1998, ApJ, 501, 1

\bibitem[Hughes (2000)]{hughes} Hughes, J. P. 2000, ApJ, 545, L53

\bibitem[Hughes, Rakowski, \& Decourchelle(2000)]{hughes00} Hughes, J. P., Rakowski, C. E., \& Decourchelle, A. 2000, ApJ, 543, L61

\bibitem[Hughes et al.(2003)]{hughl71} Hughes, J. P., Ghavamian, P., Rakowski, C. E., \& Slane, P. O. 2003, ApJ, 582, L95

\bibitem[Hwang, Hughes, \& Petre(1998)]{hwang} Hwang, U., Hughes, J. P., \& Petre, R. 1998, ApJ, 497, 833

\bibitem[Hwang et al.(2002)]{hwang02} Hwang, U., Decourchelle, A., Holt, S. S., \& Petre, R. 2002, ApJ, 581, 1101

\bibitem[Iwamoto et al.(1999)]{iwam} Iwamoto, K., Brachwitz, F., Nomoto, K., Kishimoto, N., Umeda, H., Hix, W. R., \& Thielemann, F.-K. 1999, ApJS, 125, 439

\bibitem[Koyama et al.(1995)]{koy} Koyama, K., Petre, R., Gotthelf, E. V., Hwang, U., Matsuura, M., Ozaki, M., \& Holt, S. S. 1995, Nature, 378, 255

\bibitem[Koyama et al.(1997)]{koy2} Koyama, K., Kinugasa, K., Matsuzaki, K., Nishiuchi, M., Sugizaki, M., Torii, K., Yamauchi, S., \& Aschenbach, B. 1997, PASJ, 49, L7

\bibitem[Lide(1995)]{crc} Lide, D. R., ed. 1995, CRC Handbook of Chemistry and Physics (76th ed., New York: CRC Press)

\bibitem[Long, Helfand, \& Grabelsky(1981)]{long} Long, K. S., Helfand, D. J., \& Grabelsky, D. A. 1981, ApJ, 248, 925

\bibitem[Longair(1994)]{longair} Longair, M. S. 1994, High Energy Astrophysics, Vol. 2 (2d ed.; Cambridge: Cambridge Univ. Press)

\bibitem[Mathewson et al.(1983)]{mathewson} Mathewson, D. S., Ford, V. L., Dopita, M. A., Tuohy, I. R., Long, K. S., \& Helfand, D. J. 1983, ApJS, 51, 345

\bibitem[Perlmutter et al.(1999)]{perl} Perlmutter, S. et al. 1999, ApJ, 517, 565

\bibitem[Press et al.(1985)]{press} Press, W. H., Flannery, B. P., Teukolsky, S. A., \& Vetterling, W. T. 1985, Numerical Recipes, Ed. 1 (Cambridge: Cambridge University Press), 289

\bibitem[Rakowski et al.(2003)]{rak} Rakowski, C. E., Ghavamian, P., \& Hughes, J. P. 2003, ApJ, 590, 846

\bibitem[Reynolds \& Keohane(1999)]{reykeo} Reynolds, S. P. \& Keohane, J. W. 1999, ApJ, 525, 368

\bibitem[Riess, Press, \& Kirshner(1996)]{rpk} Riess, A. G., Press, W. H., \& Kirshner, R. P. 1996, ApJ, 473, 88 

\bibitem[Riess et al.(1998)]{riess} Riess, A. G. et al. 1998, AJ, 116, 1009

\bibitem[Slane et al.(2001)]{slane} Slane, P., Hughes, J. P., Edgar, R. J., Plucinsky, P. P., Miyata, E., Tsunemi, H., \& Aschenbach, B. 2001, ApJ, 548, 814

\bibitem[Smith et al.(1991)]{smith91} Smith, R. C., Kirshner, R. P., Blair, W. P., \& Winkler, P. F. 1991, ApJ, 375, 652

\bibitem[Smith, Raymond, \& Laming(1994)]{smith94} Smith, R. C., Raymond, J. C., \& Laming, J. M. 1994, ApJ, 420, 286

\bibitem[Thielemann et al.(1991)]{thiele} Thielemann, F.-K., Hashimoto, M. A., Nomoto, K., \& Yokoi, K. 1991, in Supernovae, ed. S. E. Woosley (New York: Springer-Verlag), 609

\bibitem[Townsley et al.(2000)]{townsley} Townsley, L. K., Broos, P. S., Garmire, G. P., \& Nousek, J. A. 2000, ApJ, 534, L139

\bibitem[Truelove \& McKee(1999)]{tm} Truelove, J. K. \& McKee, C. F. 1999, ApJS, 120, 299

\bibitem[Tuohy et al.(1982)]{tuohy} Tuohy, I. R., Dopita, M. A., Mathewson, D. S., Long, K. S., \& Helfand, D. J. 1982, ApJ, 261, 473

\bibitem[Vancura, Gorenstein, \& Hughes(1995)]{van} Vancura, O., Gorenstein, P., \& Hughes, J. P. 1995, ApJ, 441, 680

\bibitem[Wang \& Chevalier(2001)]{wc} Wang, C.-Y. \& Chevalier, R. A. 2001, ApJ, 549, 1119

\bibitem[Wu et al.(1993)]{wu} Wu, C.-C., Crenshaw, D. M., Fesen, R. A., Hamilton, A. J. S., \& Sarazin, C. L. 1993, ApJ, 416, 247

\end{thebibliography}
\end{document}